\documentclass{ws-p8-50x6-00}
\usepackage{texdraw}

\begin{document}

\title{Quark-Hadron Duality}

\author{Sabine Jeschonnek}

\address{The Ohio State University, Physics Department, Lima, OH 45804
 USA\\
E-mail: jeschonnek.1@osu.edu}

\author{J. W. Van Orden}

\address{Jefferson Lab, Newport News, VA and Old Dominion
University, Norfolk, VA, USA}


\maketitle

\abstracts{Quark-hadron duality and its potential applications are
discussed. We focus on theoretical efforts to model duality.}

\section{What is Duality?}

In general, duality implies a situation in which two different
languages give an accurate description of Nature.  While one may
be more convenient than the other in certain situations, both are
correct. If we are interested in hadronic reactions, the two
relevant pictures are the quark-gluon picture and the hadronic
picture. In principle, we can describe any hadronic reaction in
terms of quarks and gluons, by solving Quantum Chromodynamics
(QCD). While this statement is obvious, it rarely has practical
value, since in most cases we can neither perform nor interpret a
full QCD calculation. In general, we also cannot perform a
complete hadronic calculation. We will refer to the statement
that, if one could perform and interpret the calculations, it
would not matter at all which set of states - hadronic states or
quark-gluon states - was used, as "degrees of freedom" duality.

However, there are cases where another, more practical form of
duality applies: for some reactions, in a certain kinematic
regime, properly averaged hadronic observables can be described by
perturbative QCD (pQCD). This statement is much more practical
than the "degrees of freedom" duality introduced above. In
contrast to full QCD, pQCD calculations can be performed, and in
this way, duality can be exploited and applied to many different
reactions.

Duality in the latter form was first found by Bloom and Gilman in
1970 in inclusive, inelastic electron scattering \cite{bgduality}.
Duality in this reaction is therefore commonly referred to as
Bloom-Gilman duality. Recently, it was impressively confirmed to
high accuracy in measurements carried out at Jefferson Lab
\cite{jlab}.

Duality also appears in the semileptonic decay of heavy quarks
\cite{isgurwise,rich}, in the reaction $e^+ e^- \rightarrow
hadrons$ \cite{weinberg}, in dilepton production in heavy ion
reactions \cite{ralf2}, and in hadronic decays of the $\tau$
lepton \cite{shifman}.

Why should one be interested in duality? It is not only a rather
interesting and surprising phenomenon, but also has many promising
applications, e.g. for experiments probing the valence structure
of the nucleon \cite{currentexps}. In inclusive electron
scattering, duality establishes a connection between the resonance
region and the deep inelastic region. Measurements in the
resonance region have higher count rates than measurements in the
deep inelastic region, and duality might be able to open up
previously inaccessible regions. New duality experiments have been
completed or are currently carried out at Jefferson Lab
\cite{exptalksthisconf,currentexps}, and there will be a large
duality program at the 12 GeV upgrade of CEBAF \cite{12gevwp}.
Examples of duality and possible applications of duality will be
discussed in the next section.

In order to use duality confidently to extract information from
experimental data, a good understanding of duality is necessary.
We need to know where it holds and how accurate it is. Our current
understanding of duality is still limited. The theoretical efforts
focus on modelling duality, and are discussed in Section 3.

\section{Applications and Examples}

Our main focus is duality in electron scattering. We will briefly
review duality in other reactions, before turning to the main
subject of the talk.

\subsection{Duality in various reactions}

For semileptonic decays of heavy quarks, duality implies that the
decay rate for hadrons is determined by the decay rate of the
underlying quark decay. For perfect duality, the quark decay rate
$ b \rightarrow c l \bar{\nu}_l$  is equal to the sum over all
hadronic decays, $\bar{B} \rightarrow X_c l \bar{\nu_l}$, where
$X_c$ stands for the ground state $D$ meson and its excited
states. For infinitely heavy masses of the $b$ and $c$ quarks,
duality was shown to hold exactly. In the realistic case of heavy,
but not infinitely heavy quarks $M_q$, the various observables
pick up correction terms of order $\frac{1}{M_q}$ and
$\frac{1}{M_q^2}$. The precise form of the correction, in
particular the question if $\frac{1}{M_q}$ corrections exist at
all, was the subject of much debate in the literature, see e.g.
\cite{isgurplb}. It seems that this matter was resolved recently
in \cite{rich}, where it was shown that the form of the correction
depends on the observable.

The reaction $e^+ e^- \rightarrow hadrons$ is a famous example of
duality \cite{halzenmartin,weinberg}. Using the optical theorem,
the cross section for the process can be described as the
imaginary part of the forward elastic scattering amplitude, where
the latter either contains a sum over all possible hadrons, i.e.
the vector mesons, or a sum over quark loops, including
interactions with hard and soft gluons. For high enough center of
mass energies, the ratio of the hadronic cross section to the $e^+
e^- \rightarrow \mu^+ \mu^-$ cross section is equal to $N_c \sum_q
e_q^2$, where $N_c$ is the number of colors and $e_q$ is the
electric charge of a quark of flavor $q$. This clearly shows that
at these energies, one may use either the quark degrees of freedom
or hadronic degrees of freedom. More surprising, even at low
center of mass energies where the resonance bumps can be seen
clearly, the "quark result" $N_c \sum_q e_q^2$, describes the
average of the ratio.

The production of dilepton pairs is the inverse reaction to $e^+
e^- \rightarrow hadrons$. One prominent feature of the dilepton
production rate is the broadening of the resonance peaks in the
spectrum, which gave rise to the explanation that the vector meson
masses drop in the medium "dropping $\rho$-mass". This phenomenon
has also been interpreted as a quark-hadron duality induced effect
\cite{ralf1,ralf2}. Just as the resonance peaks vanish for masses
larger than 1.5 GeV and below the $J/\Psi$ threshold (i. e. when
the only active flavors are $u$, $d$, and $s$), the resonance
peaks vanish in the dilepton production. Here, this vanishing
occurs at lower masses already, which was interpreted \cite{ralf2}
as an in-medium reduction of the quark-hadron duality scale of
$1.5$ GeV for active $u, d,$ and $s$ flavors. Calculations working
with hadronic degrees of freedom and quark-gluon degrees of
freedom produce the same effect.

\subsection{Duality in inclusive electron scattering}

\begin{figure}[t]
\epsfxsize=21pc 
\epsfbox{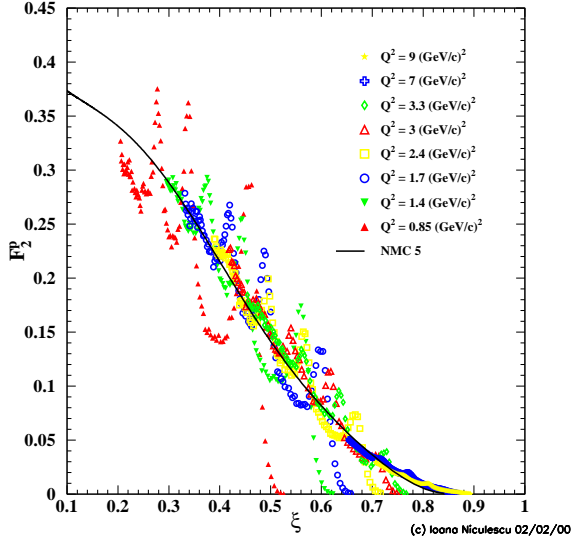} 
\caption{Experimental data for $F_2(\xi,Q^2) = \nu W_2 (\xi, Q^2)$
from Jefferson Lab \protect \cite{jlab}. The data is plotted
versus Nachtmann's variable $\xi$.} \label{figexpf2}
\end{figure}

The cross section for inclusive electron scattering is given by
$\frac{d \sigma}{d \Omega d E_f} = \sigma_{Mott} (W_2 + 2 W_1
\tan^2 {\frac{\vartheta_e}{2}}) \,,$ where $\sigma_{Mott} \propto
Q^{-4}$. Therefore, the cross section for high $Q^2$ is dropping
off rapidly. Traditionally, the region where $W$, the invariant
mass of the final state, is smaller than 2 GeV, is called the
resonance region, and $W > 2$ GeV is referred to as the deep
inelastic region. This distinction is rather artificial, and one
key point of quark-hadron duality is that these two regions are
actually connected. It is clear that duality in inclusive electron
scattering must hold in the scaling region, for $Q^2 \rightarrow
\infty$, as perturbative QCD is valid there, and therefore will
describe the hadronic reaction. In deep inelastic scattering, the
kinematics are such that the struck quark receives so much energy
over such a small space-time region that it behaves like a free
particle during the essential part of its interaction. This leads
to the compellingly simple picture that the electromagnetic cross
section in this kinematic region is determined by free
electron-quark scattering, i.e. duality is exact for this process
in the scaling region. The really interesting question is if
duality will be valid approximately at lower $Q^2$, in a region
where the cross section is dominated by resonances, which are
strongly interacting hadrons, after all. The experimental data,
see Fig.~\ref{figexpf2}, show that duality holds even at very low
$Q^2 \approx 0.5$ GeV$^2$ \cite{jlab}.
One can see clearly that the resonance data follows the scaling
curve, as given by the NMC parameterization evolved to $Q^2 = 5$
GeV $^2$. In principle, one should compare the resonance results
to the pQCD results evolved to the same $Q^2$ at which the
resonance data were taken. As the resonance $Q^2$ values are too
low for this, choosing 5 GeV$^2$ is a very reasonable approach.
Finite energy sum rules formed for the scaling (pQCD) curve and
the resonance regime further quantify the validity of duality, for
details see \cite{jlab}. Also, moments of the data have been
considered \cite{expmom}. The most striking feature of the moments
is that they flatten out at rather low $Q^2 \approx 2$ GeV~$^2$.

\begin{figure}
\epsfxsize=17pc 
\epsfbox{gmpplot.epsi} 
\epsfxsize=15pc 
\epsfbox{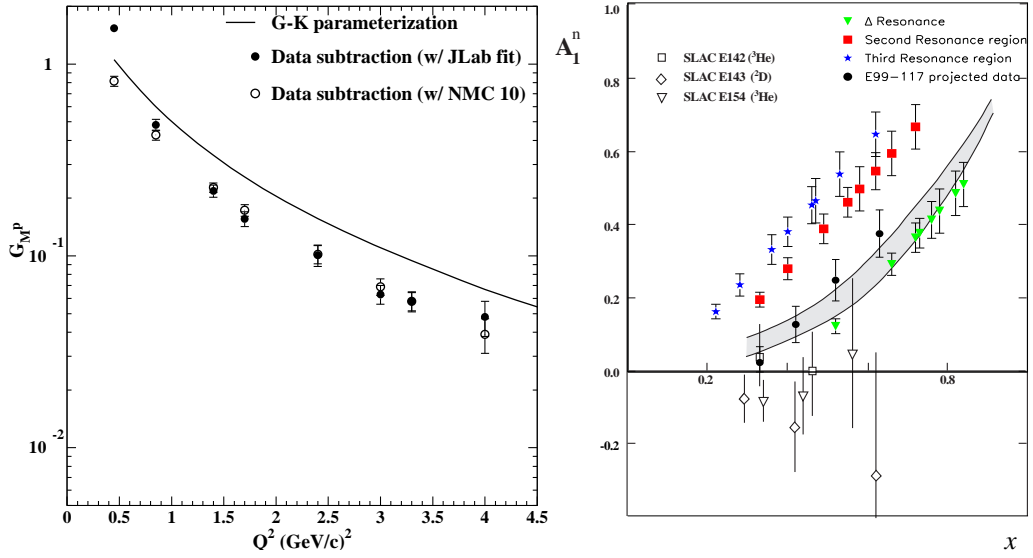} 
\caption{Left panel: Extracted values for the magnetic form factor
of the proton, from \protect \cite{expff}. Right panel: Data from
SLAC for the polarization asymmetry of the neutron, $A_1^n$, and
projected data for the Jefferson Lab experiment E01-012 \protect
\cite{nilanga}.} \label{figff}
\end{figure}

If duality holds very locally, i.e. for just one resonance,
instead of the whole resonance region, then one may use it to
extract information on the resonance region from the deep
inelastic region, and vice versa. A benchmark  for applying
duality in this, very local, way is the extraction of the magnetic
form factor of the proton from the scaling curve \cite{dgp,expff}.
The result is shown in Fig.~\ref{figff}. The qualitative agreement
is very good, and quantitatively, one sees that the duality
extraction undershoots the form factor parameterization somewhat.
This result may give us a good idea where we are in our
understanding of duality, and in our ability to extract
information from the data. One important caveat in this case is
that here, $G_M^p$ was extracted from the $F_2$ data, and a
constant ratio of 2.79 was assumed for the ratio of $G_M$ to
$G_E$. As we know from recent Hall A data from Jefferson Lab
\cite{hallaff}, this is not a decent assumption, and may introduce
a sizable error. The good news is that new data for $F_1$ have
recently been taken at Jefferson Lab, and an extraction of $G_M$
can be performed without any assumptions on the ratio $G_M/G_E$,
as in elastic scattering, the electric form factor does not enter
into the purely transverse $F_1$.

\begin{figure}
\begin{center}
\makebox[0pt]{
\input{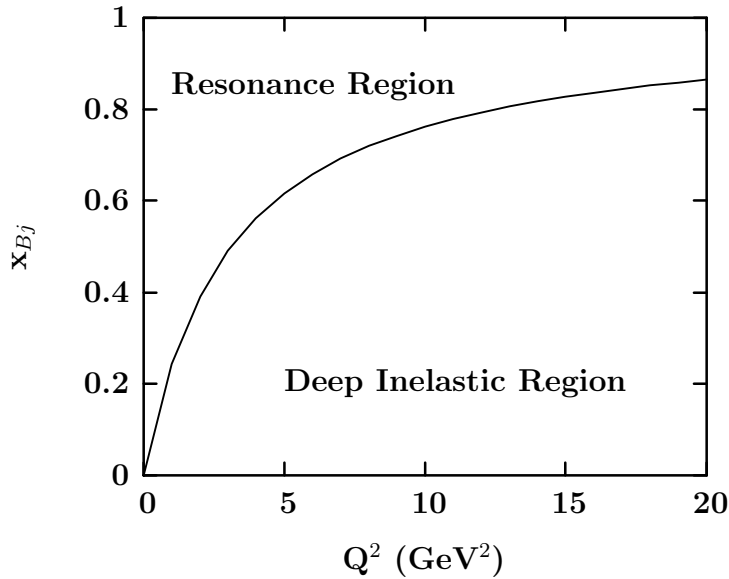}}
\\
\end{center}
\caption{The kinematic plane. The line indicates that W = 2 GeV.
The region below it corresponds to W~$>$~2~GeV (deep inelastic
region), the region above it corresponds to W~$<$~2~GeV (resonance
region).} \label{figkin}
\end{figure}

Now, while extracting $G_M$ from deep inelastic data is a good
check of our methods, this is not necessarily the "direction" we
want to take. From a practical point of view, it is very
interesting to learn about the deep inelastic region from the
resonance data. The valence quark region, i.e. the region of
$x_{Bj}$ close to 1, is of particular interest. However, data
there are scarce, as the count rate in this kinematic region is
very low. This can be seen immediately by inspecting
Fig.~\ref{figkin}, and recalling that $\sigma_{Mott} \propto
Q^{-4}$. If one wants to measure in the {\sl deep inelastic
region} at large $x_{Bj}$, one necessarily has a very large $Q^2$.
However, if one is interested in a measurement at the very same
value of $x_{Bj}$ in the {\sl resonance region}, the $Q^2$ values
may be very small, and the count rate may therefore be much
larger.

One interesting, but currently not very well known quantity is the
polarization asymmetry of the neutron, $A_1^n$. For $x_{Bj}
\rightarrow 1$, it contains information about the valence quark
spin distribution functions. There exist various, widely differing
predictions for this quantity, for a review, see \cite{nathana1n}.
If one believes that duality holds very locally, one may predict
$A_1^n$ from form factor data \cite{wallym}. As can be seen from
Fig.~\ref{figff}, the currently existing data from SLAC (open
symbols) are plagued by very large error bars, and do not reach
the region of high $x_{Bj}$ relevant for the valence quarks. The
filled symbols represent $projections$ for data that are currently
taken at Jefferson Lab, exploiting duality \cite{currentexps}.
This means that data would be taken in the resonance region, and
then properly averaged to obtain information on the deep inelastic
region. As can be seen, the data taken in this way would have a
much higher precision than the presently existing SLAC data, and
would be deep in the valence quark regime of high $x_{Bj}$.
However, before we can apply such a "duality procedure" to data
from the resonance region, we must understand where exactly
duality holds, how exact it is, and which averaging procedure
needs to be applied. The latter is especially important for
polarized measurements, as one will need to take care to average
only over resonances with the correct quantum numbers. Currently,
we do not yet have such a firm and quantitative understanding of
duality. There are several groups working on improving our
theoretical understanding of duality, which is the topic of the
next section.

\section{Search for the Origins of Duality}

The main questions that need to be addressed from the theoretical
side are: Why do we observe duality? How can we see precocious
scaling in a region where the interactions are strong? And, very
relevant to applications of duality: For which observables in
which kinematic regimes can we apply quark-hadron duality, and how
precise are our results going to be? Theoretical efforts can be
divided into two categories: refinement of the data analysis, e.g.
use of various scaling variables, Cornwall-Norton vs Nachtmann
moments, target mass corrections, averaging \cite{exp,expmom,sim},
and modelling \cite{dp1,dp2,pp,close}. I will focus on the latter
for the rest of my talk.

\begin{figure}[t]
\epsfxsize=10pc 
\epsfig{file=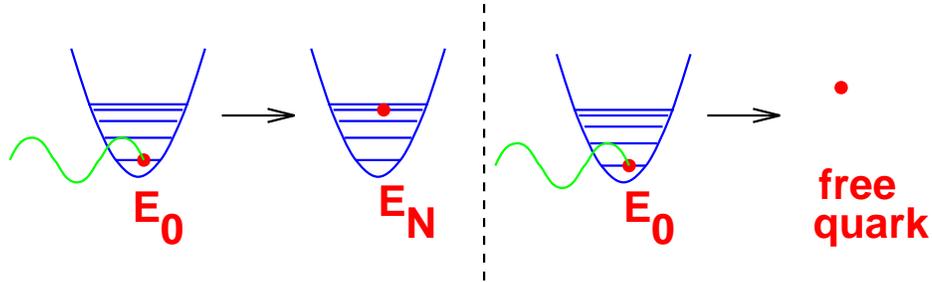, angle=270, width=12cm}
 \caption{Schematic view of the
model calculations. The left panel shows the bound-bound
transition, the right panel shows the bound-free transition.}
\label{figmodschem}
\end{figure}

When modelling duality, the first goal is to gain a qualitative
understanding of the phenomenon. Obviously, the situation as
observed in nature is very complicated, necessitating various
simplifications. Nevertheless, the goal is to incorporate the
essential physical features into a model. The general approach is
to choose a solvable model for hadrons, calculate the relevant
observables, and compare these results to the - hypothetical -
free quark results. At this point, all models assume that after
the excitation from the ground state to an excited level $N$, the
quark will remain in its excited state,i.e. the produced resonance
will not decay. The results obtained for the transition of the
quarks to a bound, excited state are summed over and compared to
the case where in the final state, the binding potential is
switched off, and the quark is "free". The latter case corresponds
to the pQCD situation. A schematic view of the modelling is given
in Fig.~\ref{figmodschem}.

All models for duality must fulfill the following criteria:
\begin{enumerate}
  \item The model must reproduce scaling. In addition, the scaling
  curve for the transition from the quark's ground state to the
  excited state must lead to the same scaling curve as the
  transition from the ground state to a free quark state.
  \item The calculated moments must flatten out for large $Q^2$, as
  observed in the data.
  \item The resonance region results should oscillate around the
  scaling curve.
\end{enumerate}

Let me now turn to one particular model, which was introduced in
\cite{dp1,dp2}. The approach in this work was to construct a model
with just a few underlying basic assumptions, which could be
extended to the more realistic case. In \cite{dp1,dp2}, we assumed
that it is sufficient to incorporate relativity and confinement in
a valence quark model. We also treated the quarks as scalars -
while spin is crucial in nature, we assumed that for duality to be
observed, it would not be necessary. In principle, we are
interested in nucleon targets, i.e. in a three-body system. As
this poses some technical difficulties, we assumed that only one
quark would carry charge and therefore interact with the photon,
the other two quarks form a spectator system. One may think of
them either as an anti-quark or as a diquark. In order to further
simplify the task, we also assumed that the spectator system has
infinite mass. This means that instead of solving the
Bethe-Salpeter equation for the two-body problem, we have to solve
only a one-body equation. As we are dealing with scalar quarks,
the Klein-Gordon equation needs to be solved. We model the
confinement using a scalar, linear potential, $V \propto r$. As
the potential enters the Klein-Gordon equation as $V^2$, the
resulting equation resembles the Schr\"odinger equation for the
non-relativistic harmonic oscillator. This has the advantage that
the wave functions obtained in the solution are exactly the wave
functions obtained for the non-relativistic harmonic oscillator,
whereas the energy spectrum is given by $E_N \propto \sqrt{N}$,
which leads to a much higher density of excited states than in the
non-relativistic case, where $E_N^{non-rel} \propto N$. A
comparison between the relativistic and the non-relativistic
solutions is easily feasible in this case. A nice feature of this
model is that the solutions can be obtained analytically. The two
parameters needed for this model are the constituent quark mass,
$m = 0.33 $ GeV and the string tension, which takes a value of
$0.16 $ GeV$^2$. None of the results depend crucially on these
precise values, and we have checked that variation of these values
gives reasonable results, e.g. we obtain the free case in the
limit of the sting tension going to zero. While all particles,
including beam and exchange particles, were treated as scalars in
\cite{dp1}, only the quarks were treated as scalars in \cite{dp2}.
In the latter case, with spin $1/2$ electrons and spin $1$
photons, one deals with a conserved current.

The first requirement of duality that must be fulfilled by a model
is scaling of the bound-bound transition, i.e. scaling for the
case where only resonances are in the final state. Before
investigating the scaling behavior, i.e. the behavior for large
$Q^2$, we need to establish which quantity ought to scale, and
which scaling variable to use. Bjorken's variable $x_{Bj} =
\frac{Q^2}{2 M \nu}$ and scaling function $F_2 = \nu W_2$ are
designed for the region of $Q^2 >> M^2$. Duality was observed to
hold at much lower values of $Q^2$, where the target mass $M$ is
about as large as $Q^2$, and the constituent quark mass, which is
the relevant quantity at the considered low $Q^2$, is not
negligible compared to $Q^2$. This situation demands a different
scaling variable and scaling function. Bloom and Gilman  used the
ad hoc variable $x' = \frac{Q^2}{W^2 + Q^2}$, and later on
\cite{barbieri}, a variable that treats target mass and
constituent quark mass on the same footing was derived. It reads
$x_{cq} = \frac{1}{2 M} (\sqrt{\nu^2 + Q^2} - \nu) (1 + \sqrt{ 1 +
\frac{4 m^2}{Q^2}})$, and was derived for the case of free quarks
with a momentum distribution. When deriving a scaling variable, it
turns out that it is intimately connected to a scaling function,
which for our case (scalar quarks), reads ${\cal{S}}_{2,cq} =
|\vec q \, | W_2$. Note that all scaling variables and scaling
functions must reduce to Bjorken's variable $x_{Bj}$ and $F_2$ in
the limit of high $Q^2$.

\begin{figure}[t]
\epsfxsize=10pc 
\epsfig{file=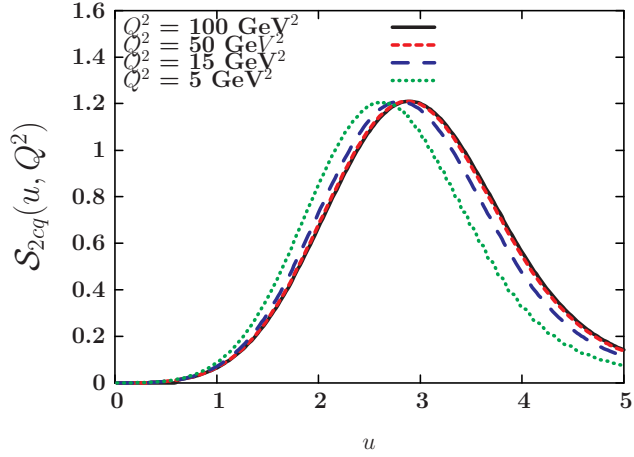,  height=6cm}
 \caption{Scaling of the bound-bound transition for $Q^2 \rightarrow \infty$.} \label{figscalbb}
\end{figure}

\begin{figure}[t]
\hspace{2cm}\epsfig{file=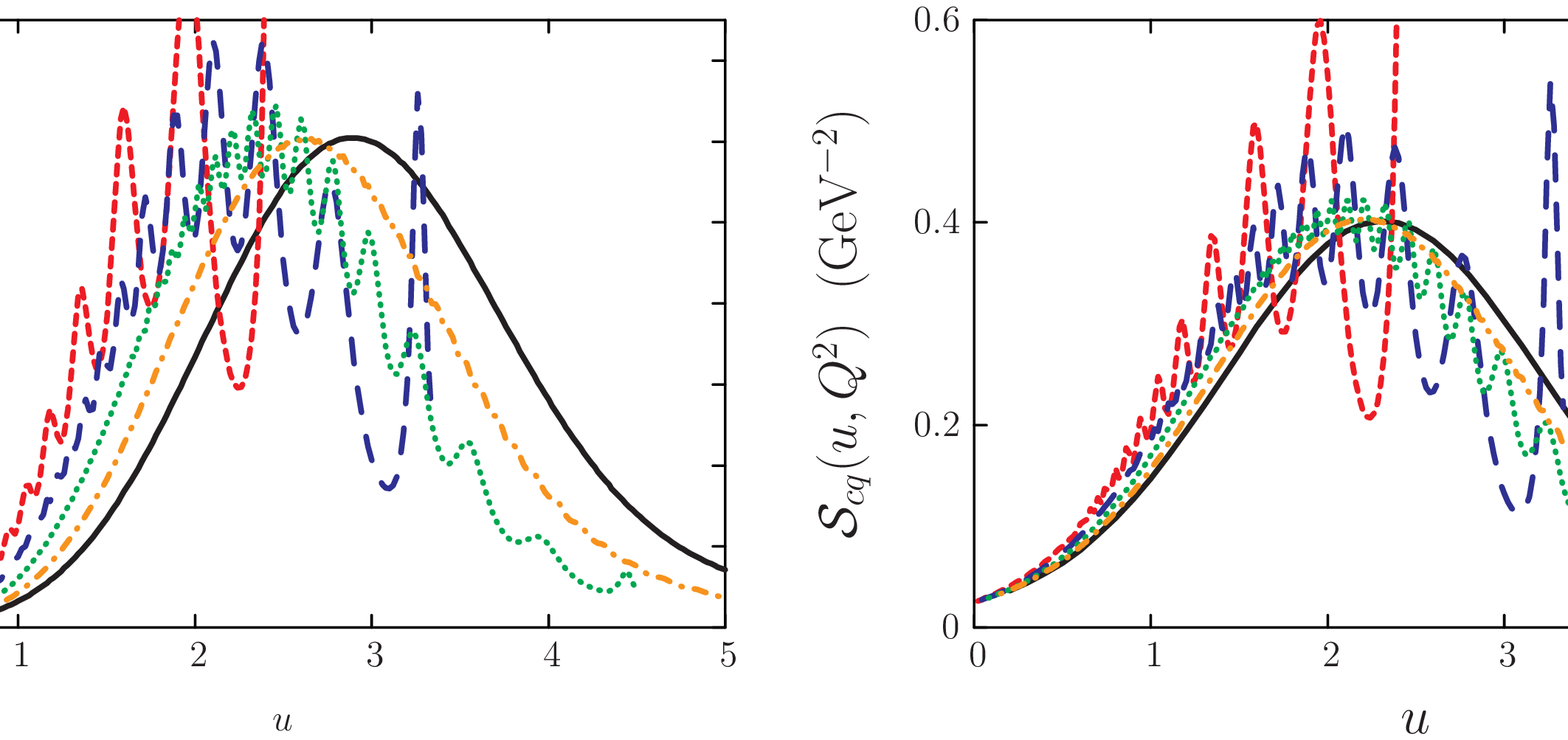, width=8cm} \caption{Duality
at low $Q^2$ for the electromagnetic current (left panel), and the
all scalar case (right panel). The solid lines show the  result
for large $Q^2$, the short dashed lines show $Q^2 = 0.5$ GeV$^2$,
the long-dashed lines show $Q^2 = 1 $ GeV$^2$, the dotted lines
show $Q^2 = 2 $ GeV$^2$, and the dash-dotted lines show $Q^2 = 5 $
GeV$^2$. } \label{figlowq}
\end{figure}

The results for the scaling in the bound-bound case are shown in
Fig.~\ref{figscalbb}. It is clear from the figure that scaling is
present: once $Q^2$ is high enough, the curves for different $Q^2$
practically coincide. Analytically, it was shown \cite{dp2} that $
{\cal{S}}_{2,cq} = \frac{m^2 u_{Bj}^2}{\pi^\frac{1}{2}\beta E_0}
\exp{\left(-
      \frac{(E_0-mu_{Bj})^2}{\beta^2}\right)} $,
and that this is the same result which one obtains for the
bound-free transition. It is interesting to note that the scaling
function obtained in the all scalar case - where again, the
bound-bound and bound-free transitions lead to the same scaling
function - has a slightly different analytic form: $
{\cal{S}}_{cq} = \frac{1} {4 \pi^\frac{1}{2}\beta E_0}
\exp{\left(-
      \frac{(E_0-mu_{Bj})^2}{\beta^2}\right)} \,.$
In the former case, one obtains that the scaling function goes to
zero for the scaling variable approaching zero, as expected for
valence quarks. However, we do not observe the behavior $\propto
\sqrt{u}$, as predicted by Regge theory \cite{robertsbook}. With
our simple model, this was not to be expected, though, and it is
interesting to observe how introducing the proper spin for the
beam and exchange particles leads to a more realistic description.
he moments flatten out at large $Q^2$, as required, and duality at
low $Q^2$ is shown in Fig.~\ref{figlowq} for the all scalar case
(right panel) and the electromagnetic case (left panel).

A similar model is discussed in \cite{pp}. These authors consider
a scalar probe and scalar quarks, and start from the
semi-relativistic Hamiltonian
 ${\cal{H}} = \sqrt{\vec p \, ^2} +
\sqrt{\sigma} r$, where the quarks are massless. The solutions
obtained in this approach are purely numerical.  When considering
scaling with respect to the many-body variable $\tilde y = \nu -
|\vec q|$, scaling and local duality are observed. The authors
also address the interesting question of contributions to sum
rules from the time-like region, which may appear due to the
binding of the quarks. The results in \cite{pp} differ in one
important aspect from the results discussed previously
\cite{dp1,dp2}: the bound-bound and bound-free transitions do not
lead to the same scaling curves, they differ by about 30 \%. This
difference apparently stems from the different wave equations used
for the two models.

\section{Summary and Outlook}

We have shown that duality appears in many reactions, is
experimentally very well established, and has interesting and
useful applications. Duality can be modelled, and with just a few
basic assumptions, one can qualitatively reproduce all the
features of duality. In the future, we will see more data
exploring duality in various reactions - unpolarized and polarized
reactions, and meson production. Theory will progress to more
realistic models, including the spin of quarks and explicitly
modelling the decay.

\section*{Acknowledgments}

We gratefully acknowledge discussions with F. Close, R. Ent, R. J.
Furnstahl, N. Isgur, C. Keppel, S. Liuti, I. Niculescu, W.
Melnitchouk, M. Paris, and R. Rapp. This work was supported in
part by funds provided by the National Science Foundation under
grant No. PHY-0139973 and by the U.S. Department of Energy (DOE)
under cooperative research agreement No. DE-AC05-84ER40150.

\end{document}